\documentclass[floatfix,twocolumn,aps,prd,showpacs,amsmath,amssymb]{revtex4}
\usepackage{bm}
\usepackage{epsfig}
\usepackage{psfrag}
\begin{document}


\title{Interaction of Hawking radiation with static sources in \\
       deSitter and Schwarzschild-deSitter spacetimes}

\author{J.\ Casti\~ neiras}
\email{jcastin@ift.unesp.br} 
\author{I.\ P.\ Costa e Silva}
\email{ivanpcs@ift.unesp.br}
\author{G.\ E.\ A.\ Matsas}
\email{matsas@ift.unesp.br} 
\affiliation{Instituto de F\'\i sica Te\'orica, Universidade Estadual Paulista,
Rua Pamplona 145, 01405-900, S\~ao Paulo, SP, Brazil}
\date{\today}

\begin{abstract}
We study and look for similarities between the response rates 
$R^{\rm dS}(a_0, \Lambda)$ and $R^{\rm SdS}(a_0, \Lambda, M)$
of a static scalar source with constant proper acceleration $a_0$ interacting 
with a massless, conformally coupled Klein-Gordon field in
(i) deSitter spacetime, in the Euclidean vacuum, which describes a thermal
flux of radiation emanating from the deSitter cosmological horizon, and in 
(ii) Schwarzschild-deSitter spacetime, in the Gibbons-Hawking vacuum, which
describes thermal fluxes of radiation emanating from both the hole
and the cosmological horizons, respectively, where $\Lambda$ is the 
cosmological constant and $M$ is the black hole mass.
After performing the field quantization in each of the above spacetimes, we 
obtain the response rates at the tree level in terms 
of an infinite sum of zero-energy field modes possessing all possible angular
momentum quantum numbers. In the case of deSitter spacetime, this formula
is worked out and a closed, analytical form is obtained. 
In the case of Schwarzschild-deSitter spacetime such a closed formula 
could not be obtained, and a numerical analysis is performed. We conclude, 
in particular, that  $R^{\rm dS}(a_0, \Lambda)$ and $R^{\rm SdS}(a_0, \Lambda, M)$ 
do not coincide in general, but tend to each other when $\Lambda \to 0$ or 
$a_0 \to \infty$. Our results are also contrasted and shown to agree 
(in the proper limits) with related ones in the literature.

\end{abstract}

\maketitle

\section{Introduction}\label{intro}
It is well
known from classical electrodynamics that accelerated electric
charges radiate as seen by an inertial observer in Minkowski spacetime. 
However, according to
the equivalence principle, a uniformly accelerated charge is seen by a 
comoving observer as being static in a ``uniform gravitational field'', and
thus it is not expected to radiate. In the classical context, this
apparent paradox was worked out in some detail by Rohrlich, Fulton~\cite{Rohrlich}
and Boulware~\cite{Boulware}. 

The same problem has also been analyzed in a quantum
context \cite{HMS}, in terms of photon emission rates,
using the fact that an observer comoving with
a uniformly accelerated charge views the latter as immersed in a
Fulling-Davies-Unruh (FDU) thermal bath \cite{FD,U}. More
specifically, the interaction of the static charge (as computed by comoving
observers) with the FDU thermal bath results in the absorption
and stimulated emission of {\em zero-energy}
Rindler photons, which, although {\em unobservable}, nevertheless exactly 
account for the usual photon emission described by an inertial observer.

A particularly interesting arena to study interactions between sources
and radiation is the vicinity of black holes, 
where the presence of non-trivial classical and quantum effects offer 
a wealth of conceptual and technical challenges. In this setting, it has 
recently been shown that the response $R^{\rm Sch} (a_0,M)$ of a pointlike 
static scalar source
with proper acceleration $a_0$ outside a Schwarzschild black hole of
mass $M$ interacting with {\em massless} scalar particles of Hawking
radiation (associated with the Unruh vacuum) is exactly the same as
the response $R^{\rm M} (a_0) \equiv q^2 a_0/4\pi^2$ (in natural units, where 
$q$ is a small coupling constant) of such a source when it is uniformly 
accelerated with the same proper acceleration in the inertial vacuum of 
Minkowski spacetime or, equivalently, when it is static in the FDU thermal bath of 
the Rindler spacetime~\cite{HMS2}. This is surprising, because structureless 
static scalar
sources can only interact with zero-energy field modes: such modes
probe the global geometry of spacetime and are accordingly quite
different in Schwarzschild and Rindler spacetimes. Indeed, this
equivalence is not verified, e.g., when either (i) the Unruh vacuum is replaced by
the Hartle-Hawking vacuum~\cite{HMS2}, (ii) the black hole is endowed
with electric charge~\cite{CM} or (iii) the massless Klein-Gordon
field is replaced with electromagnetic~\cite{CHM1} or 
{\em massive} Klein-Gordon~\cite{CCSM} ones. It is hitherto unclear 
whether the equivalence found in Ref.~\cite{HMS2} is only a remarkable 
coincidence, or if there is something deeper behind it. This circumstance 
has motivated us to study whether or not the equivalence would
persist when one includes the presence of a cosmological constant, i.e., by 
replacing Schwarzschild with Schwarzschild-deSitter (SdS) spacetime and 
Minkowski with deSitter spacetime.

SdS spacetime may be viewed as describing a spherically symmetric black 
hole immersed in a universe with a positive cosmological constant $\Lambda >0$. 
It has attracted much attention lately on account of recent type Ia 
supernovae and cosmic microwave background (CMB)
observations~\cite{cosmconst} indicating that the Universe at large
scale has (approximately) flat spatial geometry and is in accelerated expansion. 
These data suggest the existence of some background form of
energy (``dark energy'') with negative pressure. The most plausible
scenarios to describe this energy include the existence of a
positive cosmological constant and quintessence fields. Although in the latter
case the energy density of the dark energy is allowed to change in 
time, in many models this variation can be neglected for
astrophysically relevant scales. In other words, when considering
objects like black holes, one can for most purposes assume the
presence of an effective, positive cosmological constant. In light of
these facts, the interest in considering SdS black holes becomes
clear.

In this paper we consider a static, pointlike source interacting with 
a conformally coupled, massless Klein-Gordon field in both deSitter and 
SdS spacetimes. Quantum field theory in deSitter spacetime has been much 
studied in the literature~\cite{deSitter}. In this case, we take
the field state to be the {\em Euclidean vacuum}~(see, e.g., \cite{BD}) 
describing a thermal bath as seen by static observers. Our
calculations in deSitter closely follow those of Higuchi~\cite{higuchi}, but 
our presentation is somewhat different. In particular, it is useful for our 
purposes to derive the response rate of our pointlike source at a 
generic position inside the cosmological radius. 
In the SdS case, there are two main technical hindrances 
in considering the quantization of the Klein-Gordon
field. The first one is the definition of what we shall call the 
{\em Gibbons-Hawking vacuum}~\cite{GH}. This state describes a situation 
in which we have 
thermal fluxes emanating from {\em both} the hole and the cosmological event
horizons. As usually, the related temperatures 
are proportional to the corresponding surface gravities 
${\kappa}_h$ and ${\kappa}_c$. 
Because in general ${\kappa}_h \neq {\kappa}_c$, there are
technical difficulties in defining such a state~\cite{GH,KW} in the
whole SdS spacetime. However, for the region between the horizons one may
devise an heuristic prescription to define it, since in realistic situations 
where black holes are formed by gravitational 
collapse in a de Sitter background, it is natural
to expect the emission of thermal radiation from both horizons 
(see~\cite{GH} for an outline and further
justification for the mentioned prescription). We shall not
dwell on these problems in this paper but simply assume that
radiation emanates from both horizons at definite temperatures. The
second technical difficulty is related with the quantization of the scalar 
field in SdS spacetime. Due to the spherical symmetry of the problem, the
corresponding Klein-Gordon equation is easily separated, but its radial 
part, except for the near extremal case~\cite{Lemos}, does not
appear to be amenable to analytical treatment. Accordingly, we shall
proceed to its numerical resolution.

The paper is organized as follows. In Section~\ref{SecI}, we
briefly review some geometrical aspects of SdS and deSitter spacetimes
which will be useful for establishing the setting for our analysis and 
fixing notation. In Section~\ref{SecII} we present the general formalism 
to quantize a massless, conformally coupled scalar field in
the backgrounds of interest. In Section~\ref{SecIII} we apply the
formalism to deSitter spacetime with the Euclidean vacuum, 
obtaining a simple, closed analytical form for the response at the tree 
level of a static source interacting with the radiation from the cosmological
horizon. 
In Section~\ref{SecIV}, we apply the formalism to SdS spacetime with the 
Gibbons-Hawking vacuum 
and express the response of the static source in terms of a sum over the normal 
mode angular momenta, which is numerically evaluated. We then 
compare the behavior of this response with the one obtained in  the deSitter case 
(and with the related one obtained in Schwarzschild spacetime with the Unruh
vacuum~\cite{HMS2}). Finally, in Section~\ref{sec:conclusions} we finish with 
some conclusions.
Throughout this paper, we adopt natural units ($c=G=\hbar=k_B=1$), the abstract 
index notation~\cite{W} and spacetime signature $(+ - - -)$.
\section{The backgrounds: \lowercase{de}Sitter and
Schwarzschild-\lowercase{de}Sitter}\label{SecI}

Before starting our central discussion, it will be useful to recall
some geometrical features of deSitter and SdS spacetimes. We shall
briefly do so here, confining ourselves to the minimum of information
necessary to our ends. For more details, see, e.g.,~\cite{GH,HE}.
\vspace{.5cm} 
{
\psfrag{geodesicas}{\footnotesize geodesic $r=0$}
\psfrag{superficies}{\footnotesize $t = {\rm const}$}
\begin{figure}[htb]
\begin{center}
\epsfig{file=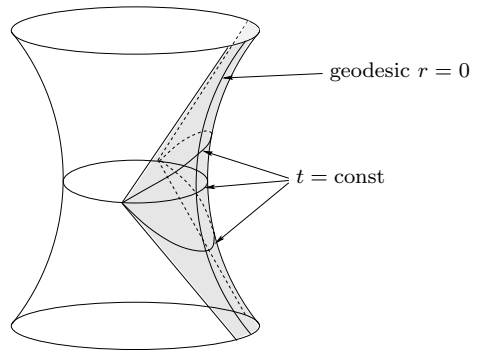,scale=.3}
\caption{\footnotesize Embedding of deSitter spacetime in a flat background
with two dimensions omitted (circular cross-sections are to be thought of as 
copies of $S^3$).  The shaded part represents the region of deSitter spacetime 
covered by the coordinates $(t,r,\theta,\phi)$. One can pick any normal, 
timelike geodesic as the origin $r=0$.}
\label{Fig01}
\end{center}
\end{figure}
}

deSitter and SdS spacetimes are vacuum solutions of Einstein's field
equations with a positive cosmological constant $\Lambda > 0$. Their 
line elements can be written as
\begin{equation}
\label{SdS}
ds^2= f(r)dt^2 - f(r)^{-1}dr^2 - r^2(d\theta ^2 + \sin ^2 \theta
d\phi^2) 
\end{equation} 
with $f(r) \mapsto f_{\rm dS}(r) = 1 - \Lambda r^2/3$ for deSitter spacetime 
and $f(r) \mapsto f_{\rm SdS}(r) = 1 - 2M/r - \Lambda r^2/3$ for SdS
spacetime~\cite{foot1}, where $M$ denotes the mass of the corresponding 
black hole. Here, the ``time coordinate'' $t$ and the
``angular coordinates'' $\theta$ and $\phi$ have their usual ranges, $
- \infty < t < + \infty$, $0 \leq \theta < \pi$, $0 \leq \phi < 2\pi$,
and for our purposes the ``radial coordinate'' $r$ must be restricted to 
non-negative values for which $f(r) > 0$.

Let us first consider deSitter spacetime. We begin by defining  
the deSitter or cosmological radius at
$\alpha \equiv \sqrt{3/ \Lambda}$, and by noting that $f(r)> 0$ implies  
$0 \leq r < \alpha$. The ``singularity'' at $r = \alpha$ is merely due to a
bad choice of coordinates, and with an appropriate
reparametrization~\cite{HE} one can obtain the corresponding maximal 
analytic extension. This spacetime has
topology $S^3 \times \mathbb{R}$ and can be isometrically embedded as
a one-sheeted hyperboloid in 5-dimensional Minkowski spacetime (see
Fig. \ref{Fig01}). The coordinates $(t,r,\theta,\phi)$ cover only part
of deSitter spacetime.
The causal structure of deSitter spacetime can be more readily
visualized through the Penrose diagram in Fig.~\ref{Fig02}. The
origin of the polar coordinates, $r=0$, and past and future
infinities ${\cal I}^{-}$ and ${\cal I}^{+}$ are represented by
vertical and horizontal borderlines, respectively.  
We note that the region labeled as I in Fig.~\ref{Fig02} [covered by 
the coordinates  $(t,r,\theta,\phi)$] on which we will focus 
has a global timelike future-directed  Killing field 
$\xi^a \equiv (\partial / \partial t)^a$. 
The Killing field $\xi^a$ becomes lightlike at  
$r = \alpha$, which comprises a {\it bifurcate Killing horizon} 
(see, e.g.,~\cite{KW} for a definition). 
The observers following integral curves of the Killing field $\xi^a$ 
in region I will be called 
{\em static} for short. Static observers have 4-velocity 
$u^a= (\xi ^c \xi _c)^{-1/2}\xi^a$, 4-acceleration 
\begin{equation}
\label{aceleracao1}
a^a = u^b\nabla _b u^a \equiv  - \frac{r}{\alpha^2} \left
(\frac{\partial}{\partial r}\right )^a,
\end{equation} 
and proper acceleration 
\begin{equation}
\label{aceleracao2}
a_{\rm dS} = \sqrt{- a^a a_a} = \frac{r}{\alpha ^2}\left (1 - \frac{r^2}{\alpha
^2}\right )^{-1/2}.  
\end{equation} 
It is thus clear that a static observer at $r=0$ follows indeed a 
geodesic. 
\vspace{.5cm} 
{
\psfrag{r=0}{\footnotesize $r=0$} 
\psfrag{r=alpha}{\footnotesize $r=\alpha$} 
\psfrag{r=infty}{\footnotesize $r=\infty$}
\begin{figure}[htb]
\begin{center}
\epsfig{file=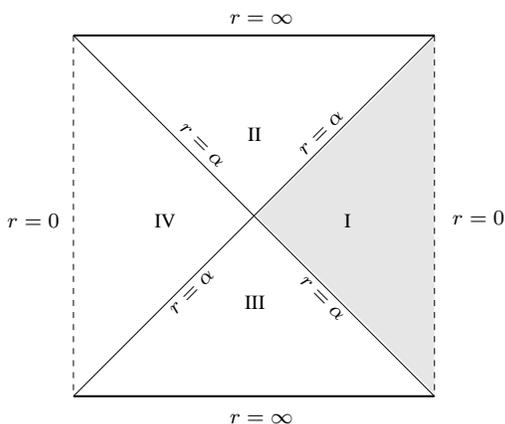,scale=.4}
\caption{\footnotesize Penrose diagram of deSitter spacetime. The
shaded region is the one covered by the coordinates
$(t,r,\theta,\phi)$. Horizontal lines cutting this diagram represent
3-spheres, and the lines labeled as $r=0$ represent the worldlines of
the ``north and south pole'' of these 3-spheres. The solid lines
labeled as $r= \infty$ correspond to past and future infinities ${\cal
I}^{-}$ and ${\cal I}^{+}$.}
\label{Fig02}
\end{center}
\end{figure}
}

Let us now turn our attention to SdS spacetime. We shall assume that  
$
{M}/{\alpha} < {1}/{\sqrt{27}}
$. 
The zeroes of $f_{\rm SdS}(r)$ are, then, found at 
\begin{eqnarray}
r_c &=& \frac{2\alpha}{\sqrt{3}}\cos \left(\frac{A}{3}\right); 
\\
r_h &=& \frac{-2\alpha}{\sqrt{3}}\cos \left(\frac{A + \pi}{3}\right); 
\\ 
r_3 &=& -(r_c + r_h),
\end{eqnarray}
where 
$ A \equiv \arccos [ - ({27 M^2}/{\alpha ^2} )^{1/2} ] $ 
satisfies
$
{\pi}/{2} < A < \pi
$.
Here, $r_c$ and $r_h$ are associated with the cosmological 
and black hole horizons, respectively, and satisfy $0< r_h < r_c$.   
Moreover, $f_{\rm SdS}(r) > 0$ for $r_h < r < r_c$.  The causal structure of the 
SdS spacetime is clear in the Penrose diagram~\cite{GH}
displayed in Fig. \ref{Fig03}. We will be interested in the 
region~I where $\xi^a = (\partial / \partial t)^a$ is a global timelike 
future-directed Killing field. 
\vspace{.5cm}
{
\psfrag{r=0}{\footnotesize $r=0$}
\psfrag{rc}{\footnotesize $r_c$}
\psfrag{rh}{\footnotesize $r_h$}
\psfrag{I}{\footnotesize I}
\psfrag{II}{\footnotesize II}
\psfrag{III}{\footnotesize III}
\psfrag{IV}{\footnotesize IV}
\psfrag{r=alpha}{\footnotesize $r=\alpha$}
\psfrag{r=infty}{\footnotesize $r=\infty$}
\begin{figure}[htb]
\begin{center}
\epsfig{file=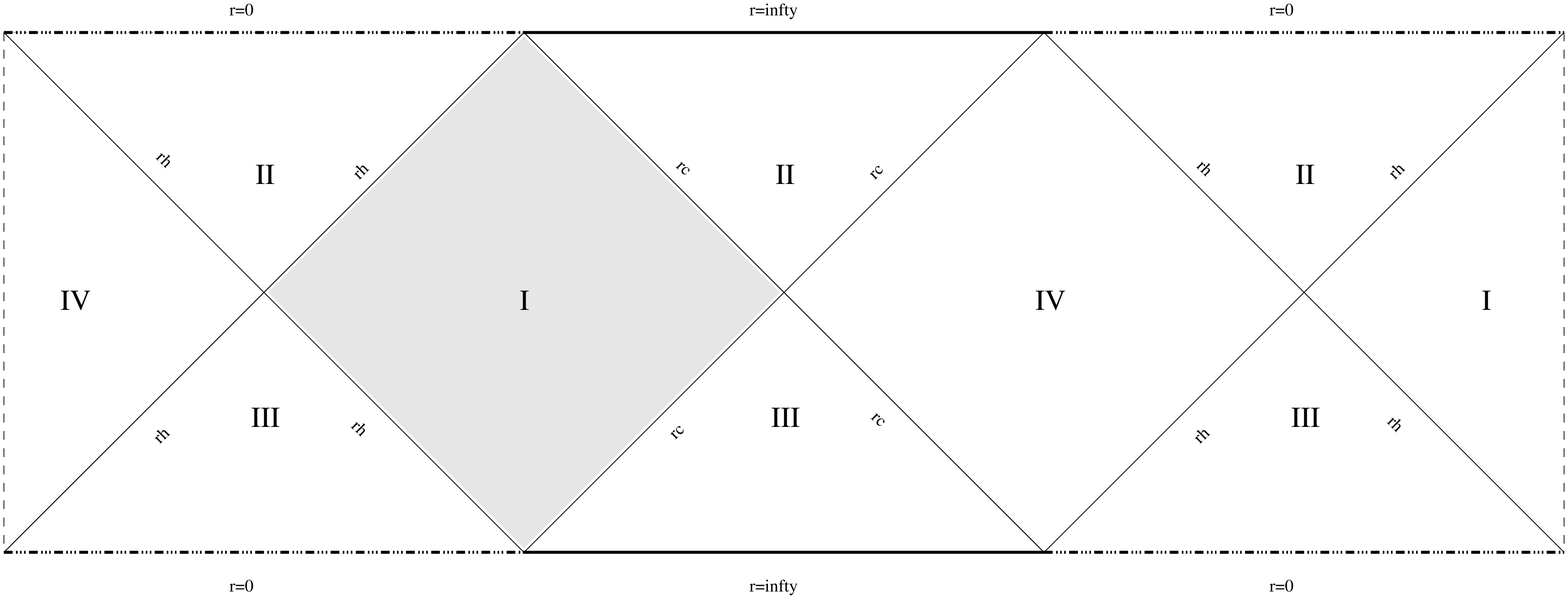,scale=.2}
\caption{\footnotesize Penrose Diagram of Schwarzschild-deSitter
spacetime. The displayed pattern repeats itself infinitely both to the
left and to the right. The shaded region is a static, globally
hyperbolic region by itself.}
\label{Fig03}
\end{center}
\end{figure}
}
In SdS spacetime, static observers
have 4-velocity $u^a= (\xi ^c \xi _c)^{-1/2}\xi^a$, 4-acceleration 
\begin{equation}
\label{aceleracaoSdS1}
a^a = u^b\nabla _b u^a \equiv  \left(\frac{M}{r^2} - \frac{r}{\alpha^2}\right) 
\left
(\frac{\partial}{\partial r}\right )^a,
\end{equation} 
and proper acceleration 
\begin{equation}
\label{aceleracaoSdS2}
a_{\rm SdS} = \sqrt{- a^a a_a} = \left|\frac{M}{r^2} - \frac{r}{\alpha^2} \right|
\left (1 -
\frac{2M}{r} - \frac{r^2}{\alpha^2}\right )^{-1/2}. 
\end{equation} 
Note that static observers with $r=(M\alpha^2)^{1/3}$ follow geodesics 
($a_{\rm SdS}=0$), due to a balance between the cosmic repulsion and the black 
hole attraction. 

\section{Response of a Static Source Interacting with a Scalar Field}
\label{SecII}

Consider now the quantization of a massless, conformally coupled
Klein-Gordon field $\Phi(x^{\mu})$, in the background defined by 
Eq.~(\ref{SdS}), described by the action
\begin{equation}
\label{KGaction}
S = \frac{1}{2}\int d^4 x \sqrt{-g}(\nabla ^a \Phi\nabla _a
\Phi - (1/6) R\Phi²), 
\end{equation}
where $g \equiv {\rm det} \{ g_{ab} \}$, and $R= 4\Lambda
= 12/\alpha^2$ is the scalar curvature for both deSitter and SdS spacetimes. 
The associated Klein-Gordon equation is   
\begin{equation}
\label{KGequation}
\nabla ^a \nabla _a \Phi + (1/6) R \Phi = 0\;.  
\end{equation}

It is well known that 
quantum field theory takes a relatively simple form in globally hyperbolic, 
stationary spacetimes where, in particular, a well defined notion of 
particle can be given  (see, e.g., Ref.~\cite{wald2} and references therein). 
This is the case for the shaded regions in Figs.~\ref{Fig02} ($0 \leq r < \alpha$) 
and~\ref{Fig03} ($r_h < r < r_c $). For each such region, we shall look for a 
set of positive-frequency modes 
\begin{equation}
\label{pfmodes}
u^i_{\omega lm}(x^{\mu}) = \sqrt{\frac{\omega}{\pi}} \frac{\psi^i_{\omega
l}(r)}{r}Y_{lm}(\theta , \phi)e^{- i \omega t} 
\end{equation}
associated with the timelike Killing field $ \xi^a = (\partial / \partial t)^a$,
where $\omega \geq 0$, $l \in \mathbb{Z} ^{+}$ and $m \in [-l,l]\cap
\mathbb{Z}$ are the frequency and the angular momentum quantum
numbers, respectively, and $Y_{lm}(\theta,\phi)$ are the spherical
harmonics. The factor $\sqrt{\omega / \pi}$ has been introduced for
later convenience. The radial part of Eq.~(\ref{KGequation}) then reads
 \begin{equation}
     \left[ 
     - f(r) \frac{d}{dr} \left( f(r) \frac{d}{dr} \right) + V_{\rm eff}(r) 
     \right] \psi^i_{\omega l}(r) =  \omega^2 \psi^i_{\omega l}(r) \;,
   \label{schrodinger1} 
\end{equation}
where the effective scattering potential  $V_{\rm eff}(r)$ is given by
\begin{equation}
\label{efetivo}
V_{\rm eff}(r) =  f(r) \left (\frac{1}{r}\frac{df}{dr} + \frac{l(l +
1)}{r^2} + \frac{2}{\alpha^2} \right ).
\end{equation}
Note that Eq.~(\ref{schrodinger1}) 
admits, in general, two sets of independent solutions which will be labeled 
with $i=I, II$. The $u^i_{\omega l m}(x^\mu)$ modes 
are 
assumed to be orthonormalized with respect to the Klein-Gordon inner 
product~\cite{BD}:
\begin{eqnarray}
\label{inner}
     i&\int_{\Sigma}& d\Sigma \; n^a 
     \left( 
     {u^i_{\omega l m}}^* \nabla_a u^{i'}_{\omega' l' m'} - 
     u^{i'}_{\omega' l' m'} \nabla_a {u^i_{\omega l m}}^*  
     \right) \nonumber \\ 
     &=& 
     \delta_{i i'} \delta_{l l'} 
     \delta_{m m'} \delta(\omega - \omega') \;,
     \\
     i&\int_{\Sigma}& d\Sigma \;  n^a 
     \left( 
     u^i_{\omega l m} \nabla_a u^{i'}_{\omega' l' m'} 
     - u^{i'}_{\omega' l' m'} \nabla_a u^i_{\omega l m}  
     \right) 
     = 0 \nonumber
\end{eqnarray}
where $n^a$ is the future-directed, unit vector normal to 
some fixed Cauchy surface $\Sigma$. These modes and their respective 
complex conjugates form a complete orthonormal basis of the space of 
solutions of Eq.~(\ref{KGequation}) in the regions of interest. 
As a result, we can expand the field operator as 
\begin{equation}
 \label{expansion}
     \hat{\Phi}(x^\mu) = 
        \sum_{i=I,II} \sum_{l=0}^{\infty} \sum_{m=-l}^{l}
        \int_0^{+\infty} d\omega 
        \left[ 
        u^i_{\omega l m}(x^\mu) a^i_{\omega l m} + {\rm H.c.} 
        \right] 
  \end{equation}
where $a^i_{\omega l m} $ and ${a^{i }_{\omega l m}}^\dagger $ are
annihilation and creation operators, respectively, and satisfy the usual 
commutation relations
\begin{equation}
     \left[ 
     a^i_{\omega l m}, {a^{i' }_{\omega' l' m'}}^\dagger 
     \right] = 
    \delta_{i i'} \delta_{l l'}
    \delta_{m m'}\delta(\omega - \omega')\;.
   \label{commutation}
\end{equation}
The ``Boulware'' vacuum $|0\rangle$ is defined by
$a^i_{\omega lm}|0\rangle = 0$ for every $i, \omega, l$ and
$m$. This is the state of ``no particles'' as defined by the static
observers following integral curves of $\xi^a$. 

Let us consider now a pointlike static scalar source lying at 
$(r_0, \theta_0, \varphi_0)$ described by 
\begin{equation}
     j(x^\mu) = (q/\sqrt{-h\;}) \delta(r-r_0)
     \delta(\theta-\theta_0) \delta(\varphi-\varphi_0) \;, 
\label{sourcecouple}
\end{equation}
where $h =- f^{-1} r^4 \sin^2 \theta$ is the determinant of the spatial 
metric induced on an equal $t$-time hypersurface $\Sigma$ and $q$ is 
a small coupling constant. This source is coupled to the 
Klein-Gordon field $\hat{\Phi} (x^\mu)$ via the interaction action
\begin{equation}
  \hat{S}_I = \int d^4x \sqrt{-g}\; j\; \hat{\Phi} \;.
  \label{interaction}
\end{equation}
All the calculations will be carried out at the tree level. 

The total response, i.e., combined particle emission and
absorption probabilities per unit proper time of the source, is given
by
\begin{equation}
     R \equiv  
       \sum_{i = I, II}
       \sum_{l=0}^{\infty} 
       \sum_{{\rm m} =-l}^{l} 
       \int_0^{+\infty} d\omega  R^{i}_{\omega l{\rm m}}  \;  ,
   \label{responsetotal}
\end{equation}
where
\begin{equation}
     R^{i}_{\omega l{\rm m}} \! \equiv  \tau^{-1} \! \left\{ |{{\cal
     A}^{i}_{\omega l{\rm m}}}^{\rm em}|^2 [1+ n^i (\omega)] + |{{\cal
     A}^{i}_{\omega l{\rm m}}}^{\rm abs} |^2 n^i (\omega) \right\} 
\label{responsepartial}
\end{equation}
and $\tau$ is the total proper time of the source. Here $ {{\cal
A}^{i}_{\omega l{\rm m}}}^{\rm em} \equiv \left\langle i \omega l {\rm
m} \left| \hat S_I \right| 0 \right\rangle $ and $ {{\cal
A}^{i}_{\omega l{\rm m}}}^{\rm abs} \equiv \left\langle 0 \left| \hat
S_I \right| i \omega l {\rm m} \right\rangle $ are the emission and
absorption amplitudes, respectively, of ``Boulware'' states $|i \omega
l {\rm m} \rangle$, and the $n^i$ factors depend on the field state 
chosen in each case.

Structureless static sources described by Eq.~(\ref{sourcecouple}) 
can only interact
with {\em zero-energy} modes \cite{HMS} and thus the response of the source 
in the ``Boulware'' vacuum vanishes. However, in the 
presence of a background thermal bath, the absorption and stimulated 
emission rates lead to a non-zero response.  
In order to deal with zero-energy modes,
we need a ``regulator'' to avoid the appearance of intermediate
indefinite results (for a more comprehensive discussion on the
interaction of static sources with zero-energy modes, see
Ref.~\cite{HMS}). For this purpose, we let the coupling constant $q$
oscillate with frequency $\omega_0$ by replacing $q$ with $q_{\omega_0}
\equiv \sqrt{2} q \cos(\omega_0 t)$ in Eq.~(\ref{sourcecouple}) and take the
limit $\omega_0 \to 0$ at the end of our calculations. The
factor $\sqrt{2}$ has been introduced to ensure that the time average
$\left\langle |q_{\omega_0}(t)|^2\right\rangle_t = q^2$ since the 
absorption and emission rates are functions of $q^2$. 
Another equivalent regularization procedure is discussed in~\cite{DS}. 
A straightforward calculation using 
$
\sum_{{\rm m} =-l}^{l} \left|Y_{l{\rm m}}(\theta_0,\varphi_0)\right|^2 =
(2l+1)/4\pi 
$
\cite{GR} gives
\begin{eqnarray}
\label{response}
& & R(r_0) = \lim _{\omega_0 \to 0} \sum_{i = I,II}\sum_{l=0}^{\infty} 
\frac{q^2 \omega_0 \sqrt{f(r_0)}}{4 \pi^2 r^2_0} 
\nonumber \\
& & (2l + 1) |\psi^i_{\omega_0 l}(r_0) |^2 [1 + 2n^i(\omega_0)] \;.
\end{eqnarray}
%
\section{Response Rate in deSitter Spacetime}
\label{SecIII}

We are now ready to consider the response rate of the static source in deSitter
spacetime. Taking $f(r) \mapsto f_{\rm dS}(r) \equiv (1 - r^2/\alpha^2)$,
the effective potential~(\ref{efetivo}) becomes
\begin{equation}
\label{efetivo2}
V^{\rm dS}_{\rm eff}(r) = \frac{l(l + 1)}{r^2}\left( 1 - \frac{r^2}{\alpha^2} 
\right) .
\end{equation}
We define a new coordinate $z = \alpha/r $, and use it to reexpress
Eq.~(\ref{schrodinger1}), with potential~(\ref{efetivo2}), in the
form
\begin{equation}
\label{legendre}
\left[ \frac{d}{dz}\left( (1 - z^2)\frac{d}{dz}\right) + l(l + 1) +
\frac{\alpha^2 \omega^2}{1 - z^2} \right] \psi^i_{\omega l} = 0,
\end{equation}
which is just the associated Legendre equation. It has two
sets of linearly independent solutions, $P^{i\alpha \omega}_l(z)$ and
$Q^{i\alpha \omega}_l(z)$ (cf., e.g., \cite{GR}), but only the
latter is regular at $r =0$ ($z = \infty$). Therefore we only consider
modes of the form
\begin{equation}
\label{pfmodes2}
u^{I {\rm dS}}_{\omega lm}(x^{\mu}) = 
C^I_{\omega l}\sqrt{\frac{\omega}{\pi}} \frac{Q^{i \alpha
\omega}_l(\alpha /r)}{r}Y_{lm}(\theta , \phi)e^{- i \omega t},
\end{equation}
where $C^I_{\omega l}$ are normalization constants to be fixed by
requiring that the modes be orthonormal with respect to the Klein-Gordon
inner product~(\ref{inner}). The physical behavior of the normal modes~(\ref{pfmodes2}) 
is clear: one may visualize them in the shaded region of
Fig.~\ref{Fig02} as emanating from the past
horizon, scattering off the
line $r = 0$, and being reflected back to the future horizon. Alternatively,
one may think of the modes ``in spatial terms'' as converging from a
2-sphere at $r = \alpha$ onto its center at $r =0$ and then spreading
out to $r = \alpha$ again. Of course, the modes ``converge isotropically'' 
only
for $l=0$, but ``swirl around'' as they plunge in onto $r =0$ for $l \neq 0$.

Let us now evaluate the normalization constants $C^I_{\omega l}$. 
For this purpose we  substitute the modes~(\ref{pfmodes2}) 
into Eq.~(\ref{inner}), where we choose $\Sigma$ to be the 
$t = 0$ hypersurface and we use the orthonormality of the 
spherical harmonics. We are then left with
\begin{equation}
\label{normalization}
{C^{I }_{\omega l}}^{\ast} C^I_{\omega'l}(\alpha/\pi) \sqrt{\omega \omega'} (\omega +
\omega')I_{l \omega \omega'} = \delta (\omega - \omega'),
\end{equation}
where 
\begin{equation}
\label{normalization2}
I_{l \omega \omega'} \equiv \int^1_0 \frac{dy}{1 - y^2} 
[Q^{i \alpha \omega}_l(1/y)]^{\ast}  Q^{i \alpha \omega'}_l(1/y).
\end{equation}
In order to evaluate $I_{l \omega \omega'}$,  we use formula $8.703$ of 
Ref.~\cite{GR} to write
\begin{equation}
\label{osques}
Q^{i \alpha \omega}_l(1/y) = \frac{\sqrt{\pi}\,
\Gamma(1 + l + i\alpha \omega) y\, f^l_{\alpha \omega}(y)}{2^{l + 1}\Gamma(l +
3/2)e^{\alpha \omega \pi}
}, 
\end{equation}
where we have defined 
\begin{eqnarray}
\label{osefes}
& & f^l_{\alpha \omega}(y) \equiv y^l(1 - y^2)^{i\alpha 
\omega / 2}   \nonumber \\
& & \times F \left(\frac{2 + l+ i\alpha \omega}{2}, 
                   \frac{1 + l+ i\alpha \omega}{2}; 
                   l + \frac{3}{2} ; y^2 \right).
\end{eqnarray}
Here, $F(a,b;c;x)$ denotes a hypergeometric
function. By using $f^l_{\alpha \omega}(y)$, 
Higuchi~\cite{higuchi} evaluated the integral
\begin{eqnarray}
\label{intdosefes2}
I^{(2)}_{l \omega \omega'} & = & \int^1_0 \frac{dy}{1 - y^2} y^2 
[f^l_{\alpha \omega}(y)]^{\ast}  f^l_{\alpha
\omega'}(y)
\nonumber \\   
&=& \frac{2 \pi}{\alpha} \left| \frac{\Gamma(l +
3/2)\Gamma(i \alpha \omega)}{\Gamma\left[(2 + l + i\alpha \omega)/2\right]\Gamma
\left[(1 + l+i\alpha \omega)/2\right]} \right| ^2  \nonumber \\
&\times & \delta(\omega - \omega') .   
\end{eqnarray}
We now use Eq.~(\ref{intdosefes2}) to compute $I_{l \omega \omega'}$  and
substitute the result into Eq.~(\ref{normalization}). Apart from an 
unimportant global phase, we get 
\begin{equation}
\label{normconst}
C^I_{\omega l} = \frac{}{ } \,
\frac{2^l e^{\alpha \omega \pi}
      \Gamma\left[(2 + l+ i\alpha \omega)/2 \right]
      \Gamma\left[(1 + l+ i\alpha \omega)/2 \right]}{
      \sqrt{\pi}{\omega}\Gamma(1 + l + i\alpha \omega)\Gamma(i \alpha
\omega)}.
\end{equation}

Next, we assume that the field is in the so-called ``Euclidean'' vacuum (also
known as ``Bunch-Davies'' or ``Birrell-Davies'' vacuum), which
describes a thermal bath of temperature
\begin{equation}
\label{temperaturedS}
T_{\rm dS} = {1}/(2 \pi \alpha)
\end{equation}
as measured by the inertial observer at $r =0$ 
(see, e.g.,~\cite{BD,KW,wald2} and references
therein for further properties of this state). 
As a consequence, 
$n^{I} (\omega) \equiv (e^{\omega\beta } - 1 )^{-1}$ 
with 
$\beta^{-1} \equiv T_{\rm dS}$. 

In order to compute the response~(\ref{response}), we 
recall that in deSitter spacetime the 
sum will be restricted to the set of regular modes, 
i.e. with  $i =I$. Then, we use 
$\psi^I_{\omega_0 l}(r_0) \equiv C^I_{\omega_0 l} Q^{i\alpha
\omega_0}_l(\alpha/r_0)$, where $C^I_{\omega_0 l}$  is obtained from 
Eq.~(\ref{normconst}), 
and the identity (cf. Eq. 8.332.1 in~\cite{GR}) 
$$
\mid x \Gamma(i x) \mid ^2 = {\pi x}/{\sinh (\pi x)}.
$$  
As a result, we obtain
\begin{eqnarray}
\label{responsemore}
& & R^{\rm dS}(r_0,\alpha) 
 = \frac{q^2 \alpha f(r_0)^{1/2}}{4 \pi^3 r^2_0}
\sum_{l=0}^{\infty} 2^{2l}(2l + 1) \nonumber \\
& &\times \left|\frac{\Gamma\left[(l + 2)/2 \right]
\Gamma\left[(l + 1)/2\right]}{\Gamma(l + 1)} \right| ^2 
\left| Q_l\left(\frac{\alpha}{r_0} \right) \right|^2,
\end{eqnarray}
where $Q_l(z)$ is the ordinary Legendre function. Now, we use the
doubling formula~\cite{GR}
$$
\Gamma(2x) =2^{2x -1} \pi^{-1/2} \Gamma(x) \Gamma(x + 1/2)
$$
with $x \equiv (l + 1)/2$ in Eq.~(\ref{responsemore}):
\begin{equation}
\label{responsemore2}
R^{\rm dS}(r_0,\alpha) = \frac{q^2 \alpha f(r_0)^{1/2} }{ 4 \pi^2 r^2_0}
\sum_{l=0}^{\infty} (2l + 1) \left| Q_l\left(\frac{\alpha}{r_0} \right) \right|
^2 \,.
\end{equation}
Finally, we use the identity~\cite{HMS2}
\begin{equation}
     \sum_{l=0}^{\infty} \left| Q_{l}(s)\right|^2 (2l+1) = 
     \frac{1}{s^2-1} \;
   \label{r34}
\end{equation}
in Eq.~(\ref{responsemore2}) to obtain the final response as a function 
of the source's position:
\begin{equation}
\label{responsedS}
R^{\rm dS}(r_0,\alpha) = \frac{q^2}{4 \pi^2 \alpha}\left(1 -
\frac{r_0^2}{\alpha^2}\right)^{-1/2}.
\end{equation}
For $r_0 = 0$ we recover the
formula for the response of an inertial source in deSitter spacetime, 
given in Ref.~\cite{higuchi}. It is convenient to invert Eq.~(\ref{aceleracao2}) 
to write the
response in terms of the source's proper acceleration, which is 
a coordinate-independent observable in General
Relativity:
\begin{equation}
\label{responsedS2}
R^{\rm dS}(a_0,\alpha) = \frac{q^2}{4 \pi^2 \alpha}(1 + \alpha^2 a_0^2)^{1/2}.
\end{equation}
We note that when $ \alpha a_0 \gg 1 $, i.e. 
when either the source approaches the cosmological horizon
or the cosmological constant is small enough ($\alpha$ being
accordingly large), we have
\begin{equation}
\label{responsedS3}
R^{\rm dS}(a_0,\alpha) \approx \frac{q^2 a_0}{4 \pi^2}.
\end{equation}
The right-hand side of Eq.~(\ref{responsedS3}) is independent of $\alpha$
and acquires the same form as the response of a static source in 
the Rindler wedge (i.e., uniformly accelerated in Minkowski spacetime) 
interacting with a massless Klein-Gordon field in the usual inertial vacuum. 
To see why this occurs, let us first write from Eq.~(\ref{aceleracao2})
the source's radial position as 
\begin{equation}
\label{approx2}
r_0 (a_0, \alpha)= \alpha [1 + (\alpha a_0)^{-2}]^{-1/2}\,.
\end{equation}
Thus, for $\alpha a_0 \gg 1$, we obtain $r_0 \approx \alpha$.
Now, in this region, the deSitter line element~(\ref{SdS}) 
reduces (apart from the angular piece) to the Rindler form 
\begin{equation}
\label{isometry2}
ds^2_{\rm dS} \approx e^{2\eta/\alpha}dt^2 - e^{2\eta/\alpha} d\eta^2\, , 
\end{equation}  
where 
$
\eta \equiv \alpha \ln (1 - r^2/\alpha^2)^{1/2}
$
($-\infty < \eta \leq 0$). 
Indeed, the {\em proper distance} between $r_0$ and the cosmological horizon 
is 
$\alpha \arctan \left[1/(\alpha a_0)\right] 
\stackrel{\alpha a_0 \to \infty}{\longrightarrow} 1/a_0$, which is 
precisely the proper distance between a static source in the 
Rindler wedge with proper acceleration $a_0$ and its horizon. 
This observation combined with the fact that the local 
temperature at the source (obtained by multiplying 
temperature~(\ref{temperaturedS}) by the Tolman factor~\cite{To}) 
corresponds to the temperature of the Fulling-Davies-Unruh 
thermal bath, 
\begin{equation}
\label{loctempcosm}
T^{loc}_{\rm dS} =     \frac{ T_{\rm dS}}{ \sqrt{f_{\rm dS}(r_0)} } 
             \approx   \frac{a_0 }{2\pi}\, ,
\end{equation}
clarifies Eq.~(\ref{responsedS3}).
In particular, the decrease in the temperature $T_{\rm dS} = 1/(2 \pi \alpha)$ 
[cf. Eq.~(\ref{temperaturedS})] when $\alpha$ grows large  
is perfectly compensated by the source's approaching to the horizon
[cf. Eq.~(\ref{approx2})]. 

\section{Response Rate in 
         Schwarzschild-\lowercase{de}Sitter Spacetime}
\label{SecIV}
We now turn to the field quantization in SdS spacetime and compare the 
response rate calculated in this case with the one 
obtained in the previous Section 
[see Eq.~(\ref{responsedS2})]. 
Similarly, we shall select Klein-Gordon orthonormalized modes 
of the form~(\ref{pfmodes}). Eq.~(\ref{schrodinger1}) with the
effective scattering potential for SdS spacetime  
\begin{equation}
\label{efetivoSdS}
V^{\rm SdS}_{ \rm eff}(r) = \left (1 - \frac{2M}{r}
- \frac{r^2}{\alpha ^2} \right ) \left (\frac{2M}{r^3} 
+ \frac{l(l + 1)}{r^2} \right )
\end{equation}
admits, now, two sets of linearly independent {\em regular} 
solutions $\psi ^i _{\omega l}$.  We shall associate 
$\psi^I _{\omega l} \equiv \psi^{\rightarrow} _{\omega l}$ 
and 
$\psi^{II} _{\omega l} \equiv \psi^{\leftarrow} _{\omega l}$, 
with purely ingoing modes emanating from the white hole horizon 
${\cal H}_h^{-}$ (hereafter simply referred as hole 
horizon) and from the past cosmological horizon ${\cal H}_c^{-}$, 
respectively. $\psi^{\rightarrow} _{\omega l}$  and 
$\psi^{\leftarrow} _{\omega l}$ are Klein-Gordon orthogonal to each other, 
as can be seen by choosing the Cauchy surface in Eq.~(\ref{inner}) to 
be $\Sigma = {\cal H}_h^{-}\cup {\cal H}_c^{-}$, and then using the fact 
that $\psi^{\rightarrow} _{\omega l}$ and 
$\psi^{\leftarrow} _{\omega l}$ vanish at ${\cal H}_c^{-}$ and
${\cal H}_h^{-}$, respectively. 

For the Gibbons-Hawking 
vacuum~\cite{GH}, the appropriate thermal factors appearing in 
Eq.~(\ref{responsepartial}) are 
$n^{I} (\omega) \equiv (e^{\omega\beta_h } - 1 )^{-1}$ 
and 
$n^{II} (\omega) \equiv (e^{\omega\beta_c } - 1 )^{-1}$, 
where
\begin{equation}
\label{temphorizons}
\beta_h^{-1} = \frac{\kappa_h}{2\pi}\;\;\; 
{\rm and}\;\;\;
\beta_c^{-1} = \frac{\kappa_c}{2\pi} \end{equation}
are the temperatures of the radiation from the hole and the
cosmological horizons, respectively, with
\begin{eqnarray}
\label{surfacegrav2}
\kappa_h  = \left. \frac{1}{2}\frac{df_{\rm SdS}}{dr}\right|_{r= r_h} 
= \frac{1}{2 \alpha^2r_h}(r_c - r_h)(r_h + \bar r) \\ 
\label{surfacegrav}
\kappa_c  = \left. \frac{1}{2}\frac{df_{\rm SdS}}{dr}\right|_{r= r_c} 
= \frac{1}{2 \alpha^2r_c}(r_c - r_h)(r_c + \bar r)
\end{eqnarray}
being the surface gravities of the corresponding horizons, 
where $\bar r \equiv r_h + r_c$. 
In this case, the response~(\ref{response}) 
can be written as
\begin{eqnarray}
\label{responseSdS}
R^{\rm SdS}(r_0,M,\alpha) 
& = & \frac{q^2 \sqrt{f(r_0)}}{2 \pi r_0^2}  
\lim_{\omega_0 \to 0} \sum_{i = I,II} \sum_{l = 0}^{\infty} (2l + 1)
\nonumber \\
&  \times & \omega_0  \mid \psi^i_{\omega_0 l}(r_0) \mid^2 n^i(\omega_0 )\,.
\end{eqnarray} 
Eq.~(\ref{responseSdS}) allows one to compute the response
provided one has obtained the (normalized) modes. For numerical convenience, 
let us introduce a new coordinate:
\begin{eqnarray}
\label{newvariable}
x(r) = 
&-& \frac{1}{2 \kappa_c}\ln \left (1 - \frac{r}{r_c}\right ) +
\frac{1}{2 \kappa_h}\ln \left (\frac{r}{r_h} - 1 \right )  \nonumber \\
&+& \frac{1}{2 \bar \kappa}\ln \left (\frac{r}{\bar r} + 1\right ),
\end{eqnarray}
where 
$
\bar \kappa \equiv (2 \alpha^2 \bar r)^{-1} (r_c + \bar r)(r_h + \bar r)
$.
In terms of the new variable $x$, Eq.~(\ref{schrodinger1}) is 
recast in the ``Schr\"odinger-like'' form 
\begin{equation}
\label{schrodinger2}
\left [ - \frac{d^2}{dx^2} + V^{\rm SdS}_{{\rm eff}}[r(x)]\right ]
\psi^i_{\omega l}(x) =
\omega^2\psi^i_{\omega l}(x).  
\end{equation}
Near the horizons [and assuming the realistic case $M \ll \alpha $ 
(i.e., $r_h \ll r_c$ )], we have
\begin{equation}
\label{asymptotic}
 V^{\rm SdS}_{{\rm eff}}(x) \sim \left \{ \begin{array}{ll} e^{-2 \kappa _c x} 
\ll 1 \;\; {\rm for}  & x \gg \alpha 
\\ e^{2 \kappa _h x} \ll 1 \;\;\; {\rm for} & x < 0, \mid x \mid \gg 2M
\end{array}
\right. .
\end{equation}
Thus, in these regions the potential becomes exponentially suppressed, and 
we can approximate Eq.~(\ref{schrodinger2}) by 
\begin{equation}
\label{schrodinger3}
- \frac{d^2 \psi^i_{\omega l} }{dx^2} \approx \omega^2\psi^i_{\omega l}(x)\,.
\end{equation}
This leads to the asymptotic behavior of the modes
\begin{equation}
\label{ingoingwh}
\psi^{\rightarrow} _{\omega l}(x) \approx \left \{ \begin{array}{ll}
A_{\omega l}(e^{i\omega x} + {\cal R}^{\rightarrow}_{\omega l}e^{- i\omega
x}) & (x < 0, \mid x \mid \gg 2M),\\ A_{\omega l}{\cal
T}^{\rightarrow}_{\omega l}e^{i\omega x} & (x \gg \alpha),
\end{array}
\right.  
\end{equation} 
and
\begin{equation}
\label{ingoingch}
\psi^{\leftarrow} _{\omega l}(x) \approx \left \{ \begin{array}{ll}
B_{\omega l}{\cal T}^{\leftarrow}_{\omega l}e^{- i\omega x} & (x < 0, \mid
x \mid \gg 2M),\\ B_{\omega l}(e^{-i\omega x} + {\cal
R}^{\leftarrow}_{\omega l}e^{ i\omega x}) & (x \gg \alpha).
\end{array}
\right.  
\end{equation}
Here, 
$\mid {\cal R}^{\rightarrow}_{\omega l}\mid ^2$, 
$\mid {\cal R}^{\leftarrow}_{\omega l}\mid^2$ 
and 
$\mid {\cal T}^{\rightarrow}_{\omega l}\mid ^2$, 
$\mid {\cal T}^{\leftarrow}_{\omega l} \mid ^2$ 
are reflection and transmission coefficients, respectively, 
for the ``scattering problem'' defined by Eq.~(\ref{schrodinger2}). 
They satisfy usual ``conservation of probability'' laws: 
$\mid {\cal R}^{\rightarrow}_{\omega l}\mid ^2 + 
\mid {\cal T}^{\rightarrow}_{\omega l}\mid ^2 = 1 $ 
and 
$\mid {\cal R}^{\leftarrow}_{\omega l}\mid^2 + 
 \mid {\cal T}^{\leftarrow}_{\omega l} \mid ^2 = 1$. 
The normalization constants $A_{\omega l}$ and $B_{\omega l}$ 
can be obtained by imposing Klein-Gordon orthonormality
of the modes $\psi^{\rightarrow} _{\omega l}$ and
$\psi^{\leftarrow} _{\omega l}$ with respect to the Klein-Gordon 
inner product~(\ref{inner}), where  Eq.~(\ref{schrodinger2}) is used  
to transform the resulting integrals into surface terms 
(see~\cite{CM} for details). Then, by using Eqs.~(\ref{ingoingwh}) 
and~(\ref{ingoingch}), we obtain 
$A_{\omega l} = B_{\omega l} = (2 \omega)^{-1}$.

The modes $\psi^{\rightarrow} _{\omega l}$ and $\psi^{\leftarrow} _{\omega l}$,
can be obtained numerically for small $\omega$ and different $l$ 
values by evolving Eq.~(\ref{schrodinger2}) with the effective 
potential~(\ref{efetivoSdS}) and the asymptotic forms~(\ref{ingoingwh}) 
and~(\ref{ingoingch}).
The corresponding total response $R^{\rm SdS}$ can be obtained, then, 
from Eq.~(\ref{responseSdS}). We note that the larger the
value of $l$, the higher the barrier of the scattering potential
$V^{\rm SdS}_{\rm eff}(r)$~\cite{CCMV} and therefore the main contributions
come from modes with small $l$. How far we must sum over $l$
in Eq.~(\ref{responseSdS}) to obtain a satisfactory numerical result
will depend on how close to the black hole horizon the 
source lies. The closer to the horizon the further over $l$ 
we must sum. 

\vspace{.5cm} 
{ 
\begin{figure}[htb]
\begin{center}
\epsfig{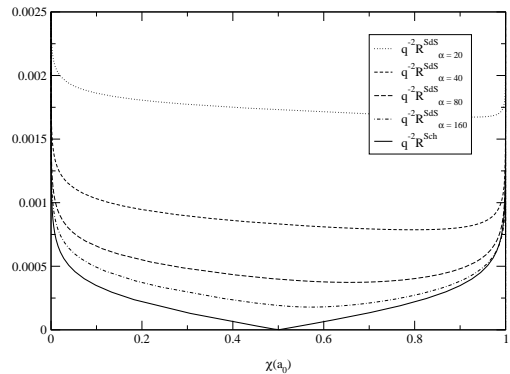}
\caption{\footnotesize $R^{\rm SdS}$ is plotted versus $\chi (a_0)$ 
(which is a monotonic function of $a_0$) for various values of $\alpha$ 
[with $M=2$, $\omega_0 = 10^{-4}$, $\rho=80$ and  the sum over $l$ is performed up to 
$l=8$ (inclusive)]. In particular, $\chi(a_0) = 0$, $\chi(a_0) = 1/2$, and 
$\chi(a_0) = 1$ correspond to 
the cases where the source is at $r_0 = r_h$, geodesic at $r_0=(M\alpha^2)^{1/3}$, 
and at $r_0= r_c$,  respectively. Note that $R^{\rm SdS}$ approaches $R^{\rm Sch}$ 
(bottom graph)  as $\alpha$ increases.
}
\label{Fig04}
\end{center}
\end{figure}
}
In Fig.~\ref{Fig04}, we plot the response $R^{\rm SdS}$ as a 
function of the source's proper acceleration $a_0$ for various 
values of $\alpha$. As originally defined, $a_0$ is not a one-to-one 
function of the source's radial coordinate $r_0$ [cf. 
Eq.~(\ref{aceleracaoSdS2})]. 
In particular, note from Eq.~(\ref{aceleracaoSdS2}) that 
$a_0 \to \infty $ either when $r \to r_h$
or $r \to r_c$. For the sake of clarity in the result 
presentation, we circumvent this feature by dropping the 
``$\mid\,.\,\mid$'''s in 
Eq.~(\ref{aceleracaoSdS2}). This is ``compensated" by plotting the 
{\em absolute value} of $R^{\rm SdS}$. After this procedure, 
the response rate is kept unchanged but the $a_0$ acquires negative sign
between the ``equilibrium" point $r_e = (M\alpha^2)^{1/3}$ 
(where $ a_0 = 0$) and $r_c$, becoming then a one-to-one function 
of $r_0$.  In special,
$a_0 \to \infty $ when $r_0 \to r_h$, but
$a_0 \to - \infty $ when $r_0 \to r_c$. Moreover, we have 
``compactified'' the range of the proper 
acceleration from $(-\infty , + \infty)$ to $(0,1)$ by 
introducing the variable 
$$
\chi(a_0) = [1 - \tanh (\rho a_0)]/2 \,,
$$
which is a monotonic function of the source's proper
acceleration $a_0$ (and where $\rho$ is a free parameter 
fixed by numerical convenience). The graph reveals that 
$R^{\rm SdS}(a_0)$ tends to $R^{\rm Sch}(a_0) = q^2 |a_0| / 4 \pi^2 $ 
 as $\alpha \to \infty $, i.e.,  
the response for a source in SdS spacetime approaches the one 
in Schwarzschild spacetime (in the Unruh vacuum) when the cosmological constant 
goes to zero (provided the source keeps the same 
proper acceleration). This result would be quite expected if we assumed that
$r_0$ (rather than $a_0$) was kept constant in the process.
This is so because when the cosmological radius $\alpha \to \infty $, 
the geometry approaches Schwarzschild's spacetime and the contribution from the
cosmological horizon in Eq.~(\ref{responseSdS}) becomes negligible 
both because $\kappa_c \to 0$ and because the low-energy modes
emanating from the cosmological horizon have to ``travel a longer
distance'' through 
the potential barrier to reach the source. As a result, these modes are
mostly scattered back, and contribute very little to the total 
response. However, the explanation is much more subtle when we consider
$\alpha \to \infty$ with $a_0$ fixed.
For $ 0 < \chi (a_0) < 1/2 $, the larger the $\alpha$ the more the 
influence of the black hole overcomes that of the cosmic expansion.
Thus, as $\alpha \to \infty$, the ``left half" of the 
curves in Fig.~\ref{Fig04} should indeed converge to Schwarzschild for the same 
reasons pointed out above. 
Nevertheless, for $1/2 < \chi(a_0) < 1$ the convergence was not expected {\it a priori} 
because in this region $r_0 > (2M\alpha^2)^{1/3} \gg 2M \approx r_h $, 
i.e., the larger the $\alpha$ the more the influence of the cosmic expansion
overcomes that of the black hole. Indeed, by neglecting the terms $M/r$ 
and $M/r^2$ in 
Eqs.~(\ref{SdS}),~(\ref{aceleracaoSdS2})~and~(\ref{efetivoSdS}), we obtain
$
f_{\rm SdS}(r)  \approx  f_{\rm dS}(r)
$,
$
|a_{\rm SdS}|   \approx  a_{\rm dS}
$
and
$
V^{\rm SdS}_{ \rm eff}(r)  \approx  V^{\rm dS}_{ \rm eff}(r)
$
(where the ``$|\;.\;|$" is used here to comply with our convention 
according to which $a_{\rm SdS} < 0$ in this region). Now, in this region,
the zero-energy modes ingoing from the black hole  
are unable to interact with the source (as confirmed by an explicit numerical 
calculation omitted here), and thus the modes ingoing from the cosmological horizon
dominate in Eq.~(\ref{responseSdS}). This indicates, at first sight, that 
$R^{\rm SdS} (a_0)$ approaches  
$R^{\rm dS}(a_0)$  [see Eq.~(\ref{responsedS2})] rather than
$R^{\rm Sch}(a_0)$. The reason why $R^{\rm SdS}(a_0)$ 
also approaches  $R^{\rm Sch}(a_0)$ in this region 
can be understood by recalling that
for large enough $\alpha$,  $R^{\rm dS}(a_0)$ approaches the response for a
uniformly accelerated source in Minkowski spacetime 
$R^{\rm M} = q^2 a_0/4\pi^2$ (see Section~\ref{SecII}),
which is in turn equivalent to $R^{\rm Sch}(a_0)$~\cite{HMS2}. 
We conclude, thus, that {\em the fact that $R^{\rm SdS} (a_0) \to R^{\rm Sch}(a_0)$ 
{\em everywhere} when $\alpha \to \infty$
is a consequence of the (non-trivial) equivalence between the responses 
in Rindler and Schwarzschild spacetimes.}

In Fig.~\ref{Fig05} we compare $R^{\rm SdS}$ with  $R^{\rm dS}$ 
[cf. Eq.~(\ref{responsedS2})], as a function of $a_0$. 
We use the same convention as in Fig.~\ref{Fig04}, but now, 
it is convenient to introduce another compactification 
variable ~\cite{note}: 
$
\zeta (a_0) = (r_0 - r_h)/(r_c - r_h),
$
where $r_0 = r_0(a_0)$ is the source's radial position as a function of the proper
acceleration obtained by inverting Eq.~(\ref{aceleracaoSdS2}) 
(without the $`` |\; . \; |''$). 
Note that, unlike $\chi$, the variable $ \zeta $ depends on both $M$ and $\alpha$, 
which are  fixed in Fig.~\ref{Fig05}. In
this figure, the various graphs correspond to the various ${\rm l}_{\rm max}$
values of the maximum $l$ used to do the sum~(\ref{responseSdS}). Note
that the sum converges very fast away from the horizons, but less so
for the regions near them. Fig.~\ref{Fig05} clearly suggests
that $R^{\rm SdS}$ coincides with $R^{\rm dS}$ near {\em both} horizons. 
That they should coincide near the cosmological horizon could be
inferred from our discussion on Fig.~\ref{Fig04}, but that 
they coincide in both horizons can be broadly understood 
from the fact that the SdS spacetime is isometric to Rindler spacetime 
near them.  This becomes manifest after the change of coordinates
$
r \mapsto \eta_h \equiv (2 \kappa_h)^{-1} \ln [f_{\rm SdS}(r)] 
$
and
$
r \mapsto \eta_c \equiv (2\kappa_c)^{-1} \ln [f_{\rm SdS}(r)] 
$,
which allows one to recast the SdS line element near the black hole and 
cosmological horizons (apart from the angular piece) as
\begin{equation}
\label{isometrySdS}
ds^2_{\rm SdS} \approx e^{2\kappa_h \eta_h}dt^2 - e^{2\kappa_h \eta_h}d\eta_h^2\,,
\end{equation}  
and
\begin{equation}
\label{isometrySdS2}
ds^2_{\rm SdS} \approx e^{2\kappa_c \eta_c}dt^2 - e^{2\kappa_c \eta_c}d\eta_c^2\,,
\end{equation}  
respectively. Now, we recall that 
the same happens for deSitter spacetime close to its horizon 
[cf. Eq.~(\ref{isometry2})]. As a result, $R^{\rm SdS}(a_0)$ turns out to 
approach $R^{\rm dS}(a_0)$ and both approach
$R^{\rm M}(a_0) = q^2 a_0 /(4\pi^2)$ [cf. Eqs.~(\ref{responsedS2}) 
and~(\ref{responsedS3})] at the horizons.


\section{Conclusions}\label{sec:conclusions}

We have quantized here a massless, conformally coupled scalar field in deSitter
and Schwarzschild-deSitter spacetimes. In both cases, the
field interacts with a static scalar source. We have computed the response 
rates of this source interacting with the 
Hawking radiation described by the Euclidean vacuum (in the deSitter case) 
and by the Gibbons-Hawking vacuum (in Schwarzschild-deSitter spacetime). 
The comparison of the responses (as functions of the proper acceleration of
the source) shows, in particular, that the equivalence obtained in \cite{HMS2}
between the responses of a static source outside a Schwarzschild black hole 
(with the Unruh vacuum) and of a uniformly accelerated source in Minkowski
spacetime (with the inertial vacuum) was not reproduced here, i.e., the
introduction of a cosmological constant breaks the original equivalence. 
\vspace{.6cm} 
{ 
\begin{figure}[htb]
\begin{center}
\epsfig{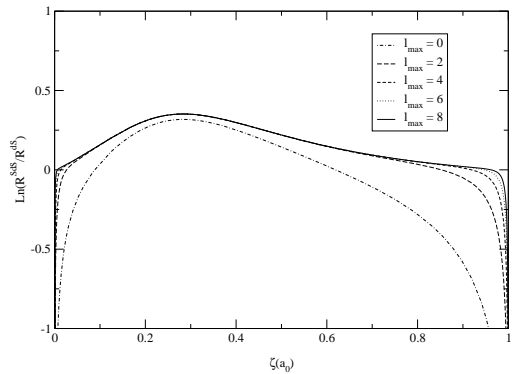}
\caption{
\footnotesize $\ln (R^{\rm SdS}/R^{\rm dS})$ 
is plotted as a function of $\zeta (a_0)$ 
(with $M=2$, $\alpha=20$ and $\omega_0 = 10^{-4}$). Here $\rm{l}_{\rm{max}}$ 
denotes the largest $l$ used in performing the
sum~(\ref{responseSdS}). $\zeta(a_0) = 0$ and $\zeta(a_0) = 1$ correspond to 
the cases where the source is at $r_0 = r_h$ and at $r_0= r_c$,  
respectively. 
}
\label{Fig05}
\end{center}
\end{figure}
}

Although the responses in deSitter and SdS spacetimes (with the respective vacua) 
do not coincide in general, very near the black hole and cosmological horizons 
they {\em are} indeed equivalent. It is so because in both these regions the source 
is expected to behave as if it were uniformly accelerated in Minkowski spacetime, 
in the inertial vacuum. We have also recovered {\em everywhere} the response of a 
static source in Schwarzschild spacetime (with the Unruh vacuum) from our response 
in SdS spacetime as the cosmological constant vanishes. We have shown that 
this is so when the source is closer to the cosmological 
horizon than to the black hole, because of the nontrivial equivalence found 
in \cite{HMS2}. 


\begin{acknowledgments} 

J.C. and I.S. would like to acknowledge full support 
from Funda\c c\~ao de Amparo \`a Pesquisa do Estado de S\~ao Paulo
(FAPESP). G.M. is thankful to FAPESP and 
Conselho Nacional de Desenvolvimento Cient\'\i fico e
Tecnol\'ogico for partial support.

\end{acknowledgments} 


\begin{thebibliography}{99}
\bibitem{foot1}
Hereafter, any subscripts (or superscripts) ``dS'' and
``SdS'' will designate quantities pertaining to deSitter and
Schwarzschild-deSitter spacetimes, respectively.
\bibitem{note} 
We emphasize that the use 
of compactification variables in this paper is entirely for 
the sake of graphical convenience. All graphs are to be thought 
of as representing functions of the source's proper acceleration.
\end{thebibliography}
\end{document}